\documentclass[aps,pre,preprint]{revtex4}
\usepackage{amsmath}
\usepackage{latexsym}
\usepackage{float}
\usepackage{amssymb}
\usepackage{graphicx}
\usepackage{epsfig}
\usepackage{graphics}

\begin{document}

\title{\bf Phase separation of an asymmetric binary
fluid mixture confined 
in a nanoscopic slit pore: Molecular-dynamics simulations}

\author{
        Katarzyna Bucior,
        Leonid Yelash,
         and Kurt Binder\\[\baselineskip]%
                 \textit{Institut f\"{u}r Physik, Johannes Gutenberg-Universit\"{a}t Mainz}\\
                 \textit {Staudinger Weg 7, D-55099 Mainz, Germany,}
}


\begin{abstract}
As a generic model system of an asymmetric binary fluid mixture,
hexadecane dissolved in carbon dioxide is considered, using a
coarse-grained bead-spring model for the short polymer, and a
simple spherical particle with Lennard-Jones interactions for the
carbon dioxide molecules. In previous work, it has been shown that
this model reproduces the real phase diagram reasonable well, and
also the initial stages of spinodal decomposition in the bulk
following a sudden expansion of the system could be studied. Using
the parallelized simulation package ESPResSo on a multiprocessor
supercomputer, phase separation of thin fluid films confined
between parallel walls that are repulsive for both types of
molecules are simulated in a rather large system (1356 $\times$
1356 $\times$ 67.8~\AA$^3$, corresponding to about 3.2 million
atoms). Following the sudden system expansion, a complicated
interplay between phase separation in the directions perpendicular
and parallel to the walls is found: in the early stages the
hexadecane molecules accumulate mostly in the center of the slit
pore, but as the coarsening of the structure in the parallel
direction proceeds, the inhomogeneity in the perpendicular
direction gets much reduced. Studying then the structure factors
and correlation functions at fixed distances from the wall, the
densities are essentially not conserved at these distances, and
hence the behavior differs strongly from spinodal decomposition in
the bulk. Some of the characteristic lengths show a nonmonotonic
variation with time, and simple coarsening described by power-law
growth is only observed if the domain sizes are much larger than
the film thickness.
\end{abstract}

\maketitle
\newpage

\section{INTRODUCTION}

Fluids confined in pores with linear dimensions on the $\mu m$ to
nm scale find increasing applications and are the subject of many
studies, both with respect to their static \cite{1,2,3,4,5} and
dynamic \cite{6,7,8,9,10,11} properties. Considering binary fluid
mixtures, it is natural to expect that the (enthalpic and
entropic) interactions between the pore walls and the fluid
particles may differ for both constituents, and then both density
and composition develop an interesting inhomogeneity in the
direction perpendicular to the pore walls. Of course, already in
the bulk the binary fluid may undergo both vapor-liquid unmixing
and fluid-fluid phase separation, resulting in complex phase
behavior \cite{12,13}. In thin slit pores, phase separation as a
thermodynamic phase transition is still possible in the lateral
directions parallel to the walls \cite{4}, and due to the possible
interplay with wetting phenomena \cite{14,15,16,17} complicated
phase diagrams are expected even for strictly symmetric mixtures
\cite{4,18}. Particularly interesting, however, is the kinetics of
these phase transitions as a function of time $t$ after a quench.
For a strictly symmetric binary Lennard-Jones mixture, where one
species is strongly attracted by the walls, it has recently been
shown by Molecular Dynamics simulations that the lateral phase
separation kinetics is characterized by a power-law for the size
of the growing domains \cite{19,20,21,22,23,24,25,26,27,28},
$\ell(t) \propto t^a$, with \cite{29,30} $a \approx 2/3$ if $\ell
(t)$ is in the range of a few Lennard-Jones diameters. While for
simple diffusive systems $a=1/3$ both in the bulk \cite{19} and in
thin films, at late enough times \cite{31}, the much faster domain
growth seen by Das et al. \cite{29,30} for a confined fluid binary
mixture may be due to some hydrodynamic mechanisms, but is not in
accord with the theoretical expectations
\cite{22,23,24,25,26,27,28}. Thus, it is interesting to study the
kinetics of phase separation for other models of confined binary
mixtures, in order to clarify which features are universal and
which features are model specific.

In the present work, we contribute to this problem by studying
phase separation for a model of a mixture of hexadecane
(C$_{16}$H$_{34})$ and carbon dioxide (CO$_2$). There are several
reasons for this particular choice: first of all, supercritical
CO$_2$ is a very important fluid in the chemical industry, useful
as a solvent in which various reactions can be carried out
\cite{32,33}, particularly applications involving polymers. Thus,
the system C$_{16}$H$_{34}$ + CO$_2$ is a prototypical polymer +
solvent system \cite{34}. Secondly, a rather simple coarse-grained
model for this system has been developed \cite{35} which describes
the experimental phase diagram rather accurately. Thirdly,
spinodal decomposition in the bulk has already been investigated
for this model by extensive simulations \cite{36}. It was found
that the system is compatible with a growth according to $\ell (t)
\propto t^{1/3}$, when $\ell(t)$ starts to exceed the
Lennard-Jones diameters, while at late times a crossover to
somewhat faster growth occurs. Limitations due to the finite
linear dimensions of the simulation box preclude strong statements
on the growth law during the late stages, however.

In Sec. 2, we shall introduce our model and briefly discuss the
simulation technique. In Sec. 3, simulation results will be
presented and discussed in the light of various theoretical
considerations. Sec. 4 contains a summary and outlook to future
work.

\section{MODEL AND SIMULATION DETAILS}
\subsection{A coarse-grained model for hexadecane + carbon
dioxide mixtures}

Although hexadecane is a very short polymer only, an all-atom
simulation of hexadecane melts would be difficult, since for a
simulation study of phase separation kinetics, length scales far
beyond the size of a molecule need to be explored, and also large
time scales are mandatory \cite{22,25,26,27,28}. Therefore, it is
advantageous to use a coarse-grained model. Coarse-graining of
polymers is usually done by taking a few chemical monomers (CH$_2$
or CH$_3$ at chain ends, in this case) together into effective
monomers, ignoring completely torsional potentials
\cite{37,38,39,40}. A successful model of this type for
C$_{16}$H$_{34}$ was proposed by Virnau et al. \cite{35},
incorporating three successive C-C bonds along a chain (plus the
corresponding hydrogen atoms) into one effective bead, so that a
chain of 5 effective monomers is created. Effective monomers along
a chain are bound together via FENE (finitely extensible nonlinear
elastic) potentials \cite{41}
\begin{equation} \label{eq1}
U_{\rm FENE} (r) = - 33.75 \varepsilon_{\rm pp} \ln [1-(r/R_{\rm
pp})^2] \quad , \quad R_{\rm pp} =1.5 \sigma_{\rm pp}
\end{equation}
\noindent where $\varepsilon_{\rm pp}$, $\sigma_{\rm pp}$ are parameters of
the Lennard-Jones (LJ) potential, that acts between all beads of
the polymer chains (bonded as well as non-bonded ones)
\begin{equation} \label{eq2}
U(r)=U_{LJ} (r) - U_{LJ} (r_{\rm cut}) \quad , \quad U_{LJ} =4
\varepsilon_{\alpha \beta} \Big[\Big(\frac{\sigma_{\alpha
\beta}}{r} \Big)^{12} - \Big(\frac{\sigma_{\alpha \beta}} {r}
\Big)^6 \Big] \quad ,
\end{equation}
\noindent where a cutoff $r_{\rm cut} = 2 r_{\rm min}$, $r_{\rm min}=2^{1/6}
\sigma_{\alpha \beta}$ is used and the potential is shifted to
zero at $r=r_{\rm cut}$ so that $U(r)$ is everywhere continuous,
with $U(r \geq r_{\rm cut})=0$. Here $\alpha, \beta=p$ (if the
particle is an effective monomer of the chains) or $\alpha,
\beta=s$ (if the particle is a solvent molecule). The parameters
$\varepsilon_{\rm pp}$, $\sigma_{\rm pp}$ and $\varepsilon_{\rm
ss}$, $\sigma_{\rm ss}$ are chosen such that the model reproduces
the experimentally known \cite{42} critical temperatures $T_c$ and
critical densities $\rho_c$ for pure C$_{16}$H$_{34}$ and pure
CO$_2$, respectively \cite{34,35}. Thus, using \cite{42} $T_c=723$K 
and $\rho_c=0.219$g/cm$^3$ has yielded \cite{34,35} (henceforth
we omit the index $p$)
\begin{equation} \label{eq3}
\varepsilon=5.79\cdot 10^{-21} J, \, \, \sigma=4.52\cdot 10^{-10} m \, ,
\end{equation}
while the experimental results for CO$_2$, $T_c=304$K and
$\rho_0=0.464$g/cm$^3$ have yielded \cite{34,35}
\begin{equation} \label{eq4}
\varepsilon_{\rm ss} =0.726 \varepsilon, \quad \sigma_{\rm ss}
=0.816 \sigma \quad .
\end{equation}
With these parameters (Eq.~(\ref{eq3}) and (\ref{eq4})) the
coexistence curves in the temperature-density plane and the vapor
pressures at coexistence as well as the interfacial tension
between the coexisting phases are reproduced in reasonable
agreement with experiment \cite{34,35}. An even better description
of CO$_2$ could be obtained by including the quadrupole-quadrupole
interaction \cite{43}, but this is out of consideration in the
present context.

The parameters $\varepsilon_{\rm ps}$, $\sigma_{\rm ps}$
for the interactions between CO$_2$ molecules and effective
monomers are described \cite{34,35} using a modified
Lorentz-Berthelot mixing rule \cite{44}
\begin{equation} \label{eq5}
\sigma_{\rm sp} =(\sigma_{\rm ss} + \sigma_{\rm pp})/2 \quad,
\quad \varepsilon_{\rm sp} = \xi\sqrt{\varepsilon_{\rm ss}
\varepsilon_{\rm pp}} \quad ,
\end{equation}
with \cite{34,35} $\xi=0.886$. While the standard
Lorentz-Berthelot mixing rule ($\xi=1$) would yield a phase
diagram topology in disagreement with the available experiments
\cite{45}, Eq.~(\ref{eq5}) gives a phase diagram in rough
agreement with these experiments \cite{34,35}. In the following,
we shall choose $\varepsilon=1$ as unit of temperature (taking
Boltzmann's constant $k_B=1$) and $\sigma=1$ as unit of length.
Fig.~\ref{fig1} shows an isothermal slice through the phase
diagram at reduced temperature $T^*\equiv k_BT/\varepsilon=1.16$ in the plane of
pressure $p$ and molar fraction $x\equiv N^s/(N^s+N^p/5)$ of carbon dioxide,  
where $N^s$ is the number of carbon dioxide molecules
and $N^p$ the number of effective monomers of hexadecane.
As will be described below, we shall simulate
pressure-jump experiments where the system suddenly is brought
from a state in the one-phase region (the initial state is
equilibrated at a density $\rho^*_{\rm tot} \equiv \rho \sigma^3 =0.8$ 
in the middle of the slit pore, which would correspond to a reduced pressure 
$p^*\equiv p \sigma^3/\varepsilon=0.34$ in the bulk system)
into the two-phase region by an isotropic increase of the volume available
to the particles.

For a system in a thin film geometry, it is also necessary to
specify the boundary conditions created by the planar walls
confining the thin film. We choose an atomistic description of
these walls, putting particles on a regular (and rigid) triangular
lattice of lattice spacing $\sigma=1$, in the $(x,y)$ plane at
$z=0.01$ and $z=0.99L_z$, $z$ being the coordinate in the direction
perpendicular to the walls. The interactions between the wall
particles $(w)$ and solvent particles or effective monomers are
described by the purely repulsive part of the LJ-potential,
Eq.~(\ref{eq2}), using $r_{\rm cut}=r_{\rm min}$ and
\begin{equation} \label{eq6}
\varepsilon_{\rm ws} = \varepsilon_{\rm wp} = \varepsilon=1, \quad
\sigma_{\rm ws}=\sigma_{\rm wp}=1 \quad .
\end{equation}
This choice was made to avoid the formation of precursors of
wetting layers of one of the species, unlike \cite{29,30}, where
an attractive interaction between the walls and one of the species
in the binary (A,B) mixture was chosen. We deliberately choose the
wall-particle interactions symmetric in the present case, to avoid
any strong preference of the wall for one of the components in our
case. However, since (unlike to \cite{29,30}) the present model is
not a symmetric mixture in the bulk, we do expect some
wall-induced concentration inhomogeneities for the present model
as well. As demonstrated recently for the case of colloid-polymer
mixtures \cite{46}, an effective attractive interaction due to
repulsive walls may arise due to purely entropic origin.

\subsection{Simulation method and preparation of the initial
state}

We study the kinetics of phase separation in thin films of CO$_2$
+ C$_{16}$H$_{34}$ mixtures by Molecular Dynamics (MD) methods
\cite{47,48,49}. As is well-known, in simple fluids and binary
fluid mixtures hydrodynamic interactions are important both for
the dynamics of fluctuations near equilibrium \cite{28,50} and for
the kinetics of coarsening in the late stages of spinodal
decomposition \cite{21,22,23,24,25,26,27,28}. MD simulations
(in the microcanonical NVE ensemble where energy $E$ is conserved
for fixed number of particles $N$ and fixed volume $V$) include
these effects of hydrodynamics implicitly and fully
\cite{47,48,49}. In fact, previous studies of phase separation in
the bulk have used this method successfully for both simple
liquid-vapor phase separation and for studies of unmixing of
binary fluid mixtures (see e.g. \cite{51,52,53,54,55}). We apply
for our system the software package ESPResSo, version 1.9.7h, 2005
\cite{56} which is particularly suitable for simulation of
coarse-grained soft matter systems on parallel computers.

For the integration of Newton's equation of motion the Velocity
Verlet Algorithm \cite{47,48,49} is applied, choosing an
integration time step $\delta t=0.002 \, \tau$, where the MD time
unit here defined as $\tau=\sigma (m/\varepsilon)^{1/2}=1$ corresponds 
to 500 integration steps.
The masses $m$ of CO$_2$ molecules, effective
monomers and wall particles are set for simplicity equal to each other, 
and time units are chosen such that $m=1$.

The initial state is created by first using a small simulation box
$L _x \times L_y \times L_z$ with $L_x = L_y=20$, $L_z=12$ 
(measured in units of $\sigma$ through this paper) with
two repulsive walls at $z=0.01$ and $z=0.99L_z$, as described in Sec.
2.1, and periodic boundary conditions in $x$ and $y$-directions.
Into this box, hexadecane molecules were inserted, having
selfavoiding walk configurations, and the CO$_2$ particles were
inserted at random position, at molar fraction of CO$_2$ $x=0.6$, such that
the initial state reaches a reduced total monomer density of 
$\rho^*_{\rm tot}=0.8$ in the center of the thin film. 
The CO$_2$ particles were only allowed to
be put at positions outside of a sphere of radius $\sigma=1$ of
each bead, to avoid that in the initial state very large repulsive
forces occur. Choosing a Langevin thermostat \cite{47,48,49}, the
system then is equilibrated at $T^*=1.16$ and is
replicated three times in $x$ and $y$-directions, to obtain a
system with linear dimensions $L_x=L_y=60$. This 9 times larger
system then is equilibrated again, for a time of 300 MD time units 
($1.5\times 10^5$ MD steps), to 
remove the effects due to original periodicity at $L_x=L_y=20$. It
was carefully tested that for the chosen conditions (i.e., for a 
supercritical solution of relatively short chains) such a short
re-equilibration time actually was enough. Then the thermostat was
switched off, and  a Galilei transformation of particle velocities
was applied to remove the motion of the center of mass of our
model system. This still rather small system, as described above,
was only used for testing our simulation and analysis procedures 
as well as for choosing optimal parameters for the pressure-jump 
simulations.
To obtain the initial state of the full system at the desired
dimensions $L_x=L_y=240$, $L_z=12$, the system was replicated
again 4 times in $x$- and $y$-direction, and the procedure of
equilibration and Galilei transformation, as described above, was
repeated again. The structure factor of the system was carefully
analyzed to check that any signs of the Bragg peaks (due to the
periodic arrangement of the replicas) have disappeared. Then the
system was equilibrated further for 400 MD time units 
($2\times 10^5$ MD steps), before the
quench was started. Note that at this stage we have already a
total number of $N=589999$ particles, namely 294000 chain segments
(i.e., 58800 chains), 88400 solvent particles and 207599 wall
particles. Since each CO$_2$ solvent particle contains 3 atoms,
and each C$_{16}$H$_{34}$ chain contains 50 atoms, the total
number of atoms (if we had an atomistic model) in our system would
be 3205200 (not counting the wall atoms).

The pressure-jump quenching experiment has the effect that the
system after the sudden quench can take a larger volume, and since
the particle numbers always are fixed, this corresponds to a
decrease of density. We do not attempt to precisely mimic how the
pressure jump is carried out in an actual experiment, but we
simply rescale the positions of the centers of mass of hexadecane
and carbon dioxide molecules in three directions such that the
final dimensions of the simulation box were $L_x$=$L_y$=300 and
$L_z=15$. Of course, one must not simply rescale all the
coordinates of the effective monomers, since
the conformation of an individual C$_{16}$H$_{34}$ molecule 
(bond lengths and positions of the monomers along a chain relative 
to each other) 
should not be
rescaled but rather stays the same, just the molecules are moved farther
apart form each other at lower density. Note that due to
Eq.~(\ref{eq3}) this final size of the box corresponds to
$L_x=L_y=1356$~\AA~while $L_z=67.8$~\AA, so the system still is a
ultrathin nanoscopic film. Wall particles were removed from the
system before the rescaling of CO$_2$ and C$_{16}$H$_{34}$
positions and inserted just after the rescaling procedure, so that
the arrangement of the wall particles (and the distances between
them) stay exactly the same as before the quench. In addition, we
reduce the energy of the system by rescaling the kinetic
energy of the particles to ensure that the temperature of the
system after the quench becomes very similar to the initial
temperature of the equilibrated homogeneous system. Physically the
walls confining a thin fluid film would be massive solid walls of
a suitable device, of course, and thermostating the walls would
have the effect of maintaining constant  temperature conditions.
Our procedure is meant as a short-cut for such a situation.

The simulation of the system after the quench is performed in the
NVE ensemble for 4000 MD time units corresponding to two million 
MD time steps. 
For the first 200 MD time units ($10^5$ MD steps), 30 runs
were performed in parallel, storing configurations after every 10
MD time units. For later times, due to the large computational
effort for our system, five systems were propagated and
configurations were analyzed for every 100 MD time units only. These
simulations were carried out on the multiprocessor system JUMP of
the John von Neumann Institute for Computing (NIC) at J\"ulich,
utilizing 16 processors in parallel, and the cluster of the
SOFTCOMP EU Network of Excellence, utilizing 4 processors in
parallel.

\section{SIMULATION RESULTS}
\subsection{Transient segregation between solvent and polymer forming a layered state}

Fig.~\ref{fig2} shows typical snapshot pictures that illustrate
the time evolution of the phase separation process in the thin
film. These snapshots show quasi-two-dimensional slices parallel
to the wall, the left column being about 3 layers (of total 10) away from the
wall, the right column being close to the center of the film (layer 5).
Already these pictures show an interesting interplay of phase
separation in the directions perpendicular and parallel to the
confining walls: in the initial stage, $t=10$ (Figs.~\ref{fig2}a, f), 
the system is still
laterally homogeneous, apart from very strongly localized density
fluctuations, but there is a strong variation of density across
the film: most of the effective monomers are concentrated in the
center of the film (Fig.~\ref{fig2}f). 
This observation is still true at $t=50$ (Figs.~\ref{fig2}b,g), but
now lateral phase separation has clearly started: in the center of 
the pore (Fig.~\ref{fig2}g),
the white ``holes'' mean that CO$_2$ bubbles with a few hexadecane molecules 
(i.e., a dilute solution of chains in supercritical CO$_2$) have formed within
the concentrated C$_{16}$H$_{34}/$CO$_2$ solution, 
while near the walls (Fig.~\ref{fig2}b) we rather have
ramified clusters of C$_{16}$H$_{34}$ molecules in a CO$_2$-rich
fluid. At later times, these structures coarsen ($t$=100, 200)
and, at the same time, the difference in density between the
center of the thin film and the regions near the walls diminishes.
For $t=1000$ (as well as for later times, that are not shown
here) the density difference has almost vanished (Figs.~\ref{fig2}e, j). 
What is more
important, the regions in the $(x,y)$-plane where the
CO$_2$-C$_{16}$H$_{34}$ interfaces occur, are identical in the
left and the right snapshot: We can picture the phase separation
in the late stages, where the characteristic linear dimensions of
the growing CO$_2$ bubbles in the concentrated C$_{16}$H$_{34}$/CO$_2$ 
solution in $x,y$-
direction are much larger than the film thickness $L_z$, simply as
a quasi-two-dimensional arrangement of flat cylinders of height
$L_z$, forming bridges between the two walls. While some of the
CO$_2$-droplets at $t=1000$ still deviate strongly from a
circular cross-section in $x,y$-directions, actually an inspection
of snapshots at still later times, such as $t=4000$ (not shown
here) shows that the droplets in fact do develop towards becoming
a circular cross section, thus, minimizing the interfacial area 
(and energy).
Ultimately (at time $t \rightarrow
\infty$ in a macroscopically large system in $x,y$-directions) we
expect a population of strictly regular cylinders of typical
radius $R(t)$ connecting the two walls and with $R(t)$ growing 
to infinity as well.
Unfortunately, for $t=4000$ in our system
the number $n(t)$ of cylinders was only $n(t=4000)=8$, implying
that the data suffer from very strong finite size effects! In
fact, studies of coarsening in simple diffusive models have
suggested that finite size effects become important already when
the number $n(t)$ of growing domains becomes distinctly smaller
than $n(t)=20$ \cite{57}, and hence our data for $t \geq 1000$
clearly suffer from finite size effects, despite the rather large
linear dimensions and number of particles in our system. 
Therefore, there is no point in
carrying out our MD simulations for the system dimensions chosen here
after longer times.

Fig.~\ref{fig3} shows now the laterally averaged total density of
particles which we define as follows
\begin{equation} \label{eq7}
\rho_{\rm tot} (z)=(N^s_i + N^p_i)/V_i \quad ,
\end{equation}
\noindent where $N^s_i$ is the number of solvent (carbon dioxide) molecules in layer
$z\equiv z_i$ with $z_i$ located in the middle of the
interval $\Delta z= 0.15$; $N^p_i$ is the number of effective 
monomers of hexadecane
in this layer,  
and $V_i=L_xL_y \Delta z$ the associated volume of layer $i$.
We have tested that the dependence of
the profile $\rho_{\rm tot}(z)$ on the width $\Delta z$ of this
volume slice $V_i$ is not important. Our choice was taken to
ensure that fluctuations in $\rho_{\rm tot}(z)$ are small enough
but no significant information on the inhomogeneity of the system
in $z$-direction is lost. One can see that near both walls there
is always a region (of thickness $\approx 0.82\sigma$) essentially free of particles,
and then the density both in the initial state and in the final
state gradually increases to an almost constant density in the
inner part of the film, 
$\rho_{\rm tot}(z/L_z\approx0.5)\approx0.8$ before the quench and 
$\rho_{\rm tot}(z/L_z\approx0.5)\approx0.4$ during the late
stages. However, at early times after the quench, most of the
particles accumulate in the center of the film, so the system
initially takes a state which shows a phase separation of the 
liquid-vapor type in the $z$-direction perpendicular to the walls:
vapor layers occur close to the walls, and a fluid (where CO$_2$
and C$_{16}$H$_{34}$ are still almost homogeneously mixed) occurs
in the center of the film. However, this vertically separated
state is unstable against lateral phase separation, in the
directions parallel to the walls, as we have already seen from the
snapshot pictures in Fig.~\ref{fig2}. 
Thus, in a particular distance $z$ from the
walls the density $\rho_{\rm tot} (z)$ approaches its final
equilibrium in a non-monotonic fashion as a function of time,
e.g., for $z/L_z =0.2$ the density decays fast from a rather large
value ($\rho_{\rm tot}\approx 0.6$) to a very small value 
($\rho_{\rm tot}\approx 0.07$) immediately after the quench, and only
when the lateral phase separation starts the density increases
again.
Note that the two-stage character of the phase separation process, where first a stratified structure forms, with low density in the film center, which then laterally decomposes, is not a consequence of studying a binary mixture, but rather a consequence of purely repulsive wall-particle interactions. In fact, we have checked for pure CO$_2$ that a similar behavior occurs.

It is also interesting to study the density profiles $\rho_s(z)
=N_i^s/V_i$ and $\rho_p(z) =N^p_i/V_i$ of the solvent particles
and the monomers separately, and to define also a profile $c(z)$
of the relative concentration of CO$_2$,
\begin{equation} \label{eq8}
c(z) = \rho_s(z)/[\rho_s (z) + \rho_p(z)] \quad .
\end{equation}
\noindent
Fig.~\ref{fig4} displays the latter profile: one can see that
despite the fact that we have chosen the same repulsive potential
between wall atoms and the monomers or solvent particles,
respectively, nevertheless the relative concentration of CO$_2$
near the walls is strongly enhanced, both in the initial state and
during the late stages of phase separation. Actually, the curves
for $c(z)$ in the initial state and in the late stages $(t \geq
500)$ almost fall on top of each other! Only in the early times
after the quench ($10 \leq t \leq 100)$ we do see a much stronger
variation of $c(z)$: near the walls almost pure CO$_2$ phase is
reached. So the phase separation clearly proceeds in two stages:
induced by the walls, first in the direction perpendicular to the
walls a layered structure forms, CO$_2$ and C$_{16}$H$_{34}$ get
almost completely segregated, with the polymer film in the center
and two CO$_2$-C$_{16}$H$_{34}$ interfaces near $z/L_z
\approx0.25$ or 0.75, respectively. However, this state costs far
too much (interfacial) energy, and is hence unstable towards
lateral phase separation. Both in the initially homogeneous state
and in the final state with the ``cylindrical'' CO$_2$-domains
across the thin film (Fig.~\ref{fig2}) we have a strong
concentration enhancement of CO$_2$ near the walls. This
enhancement clearly is an entropic effect, from the point of view
of configurational entropy polymers tend to avoid the regions
close to the walls.

Of course, for times $t < 100$ the system clearly is rather far
from equilibrium, and its state is changing rather rapidly.
Monitoring the temperature (from the kinetic energy)
\cite{47,48,49} and the  pressure (using the virial theorem
\cite{47,48,49}) as a function of time after the quench, a
distinct but relatively small increase with time in the region $10
\leq t \leq100$ is indeed found (Fig.~\ref{fig5}). However, for
later times both quantities settle down at constant values, as
desired.

\subsection{Equal-time structure factors}

In experimental studies of phase separation kinetics, most often
the equal-time structure factor $S(\vec{q},t)$ is monitored,
$\vec{q}$ being the wavevector of a scattering experiment
\cite{22,23,24,25,26,27,28}. For a thin film geometry, $\vec{q}$
needs to be oriented in the $(x,y)$-plane, of course,
$\vec{q}=\vec{q}_{||}$. Also, due to the inhomogeneity of the
system in the $z$-direction (Figs.~\ref{fig3},~\ref{fig4}), it is
of interest to distinguish in the structure factor from which
slice ($z$) the scattering particles contribute to the scattering
intensity. Moreover, having two components (which here we
symbolically denote as A and B, in order to make contact between
our notation and the relevant literature \cite{58,59}) one must
distinguish partial structure factors and those which monitor
density and concentration fluctuations. Thus, we define the
partial structure factors, resolved with respect to the
$z$-coordinate, as follows
\begin{equation} \label{eq9}
S_{\alpha \beta} (\vec{q}_{||}, z, t) = \frac{1 }{N}
\sum\limits_{k=1}^{N_\alpha} \sum\limits_{\ell=1}^{N_\beta}
\langle \exp [i \vec{q} \cdot (\vec{r} _{k \ell} (t))] \rangle,
\end{equation}
\noindent where $\alpha, \beta =A$ or $B$, 
$\vec{r}_{k \ell} (t)= \vec{r}_{k} (t) - \vec{r}_\ell (t)$, and
$N_\alpha$, $N_\beta$ being the numbers of particles of type A or
B in the slice at $z$ (i.e., the coordinates $z_k(t)$, $z_\ell
(t)$ of the particles must be in the range $z-\Delta z \leq z_k
(t)$, $z_\ell (t) \leq z + \Delta z$). While ideally one would
like to consider the limit $\Delta z \rightarrow 0$, in practice
we had to choose a rather larger value of $\Delta z$ (namely
$\Delta z =0.75$) in order to get enough statistics.

For fluids $S_{\alpha \beta} (\vec{q}_{||}, z,t)$ depends only on
the magnitude $q_{||}$ of $\vec{q}_{||}$ and not its direction.
Thus, the structure factors monitoring fluctuations of number
density $(S_{\rm nn})$ and of concentration $(S_{\rm cc})$ 
are defined as
follow,
\begin{equation} \label{eq10}
S_{\rm nn} (q_{||}, z,t) =S_{AA} (q_{||}, z,t) + 2 S_{AB} (q_{||},
z,t) + S_{BB} (q_{||}, z,t) \quad,
\end{equation}
\begin{equation} \label{eq11}
S_{\rm cc} (q_{||}, z,t) = x_B^2 S_{AA} (q_{||}, z,t) + x_A^2 S_{BB}
(q_{||}, z,t) - 2 x_A x_B S_{AB} (q_{||}, z,t) \,
\end{equation}
\noindent where $x_A=N_A/(N_A+N_B)$ and $x_B=N_B/(N_A+ N_B)$ are the
relative concentrations of A(B) particles in the slice centered at
$z$. To simplify the notation in the following, the index $||$ from
$q_{||}$ will be omitted.

Note that due to the motion of particles in the $z$-direction,
$N_A$ and $N_B$ are not conserved for a selected layer; 
in particular, during the early
stages of phase separation, $x_A$ and $x_B$ change strongly with time.
As an example, Fig.~\ref{fig6} presents $S_{\rm nn}(q,z,t)$ as a
function of wavenumber $q$ for different choices of $z$ and three
times: before the quench (a), at $t=40$ (b) and $t=100$ (c) after the quench.
The values of $z$ shown in the figure are symmetric around the center
of the film, which occurs at $(L_z/2)$, and therefore pairs of
curves should superpose, apart from statistical errors. We see
that this symmetry indeed is rather well satisfied 
(e.g., the curves for the layers 5 and 6 are indistinguishable 
from each other in the scale shown in Fig.~\ref{fig6}c), and hence the
statistical errors of our data indeed are rather well under
control. One can see a peak near $q=2 \pi$ which changes
relatively little with time: this peak and the structure at still
larger $q$ reflect the local packing of particles in a dense
fluid. Apart from the values of $z$
very close to the walls (e.g., for layers 1 and 10, 
where almost no particles occur, as the
density profiles $\rho (z,t)$ in Fig.~\ref{fig3} show), 
all curves exhibit a minimum somewhere in
the region $2 \leq q \leq 4$, while for smaller $q$, 
the structure factor $S_{\rm nn} (q,z,t)$ increases again. 
In equilibrium, the maximum of the structure factor occurs for 
$q \rightarrow 0$ (Fig.~\ref{fig6}a), as expected, 
while after the quench for large enough times, $S_{\rm nn}(q,z,t)$ 
exhibits a well-defined maximum for small $q$ (Fig.~\ref{fig6}c):
this small-angle scattering is the ``hall mark'' of spinodal
decomposition. However, close to the walls (i.e., for 
layers 2, 3, 8 and 9 centered at $z=2.25$, $3.75$, $11.25$ and
$z=12.75$, respectively, in Fig.~\ref{fig6}b) 
the scattering intensity for small $q$
does not seem to decrease again, and so the maximum is much less
pronounced. Indeed, this range of $z$ clearly exhibits a lack of
conservation of the density, due to the rapid change of the total
density $\rho(z,t)$ in this regime of times (cf. Fig.~\ref{fig3}).

The analysis of $S_{\rm cc}(q,z,t)$ gives a similar picture, and hence
is not shown here. We rather try to use both $S_{\rm nn} (q,z,t)$ and
$S_{\rm cc}(q,z,t)$ to extract characteristic lengths $R(t)$ by taking
suitable ratios of moments \cite{22}. We define $R_1(z,t)$ from
$S_{\rm nn} (q,z,t)$
\begin{equation} \label{eq12}
R_1 (z,t)= 2 \pi \sum\limits_{q=0}^{q=q_{\rm cut}} S_{\rm nn} (q,z,t)
/ \sum\limits_{q=0}^{q=q_{\rm cut}} q S_{\rm nn} (q,z,t) \quad ,
\end{equation}
\noindent and similarly for $R_2(z,t)$ from $S_{\rm cc}(q,z,t)$. 
The resulting data are shown in
Figs.~\ref{fig7} and \ref{fig8} for different choices of $z$ and
different values for the cutoff $q_{\rm cut}$.

Data for $1 \leq t \leq 10$ were not included, since at such
extremely short times after the quench both pressure and
temperature still are rather strongly time-dependent, the system
is very far from equilibrium in all respects, and a discussion of
the evolution of the system in terms of the concepts on coarsening
\cite{22,23,24,25,26,27,28} would be rather misleading. Also the
behavior in the next decade, $10 \leq t \leq 100$, is difficult to
interpret: we see an unusually strong dependence of both
$R_1(z,t)$ and $R_2(z,t)$ on the cutoff $q_{\rm cut}$, and for
some values of $z$ there occurs a slight maximum at about $30 \leq
t \leq 60$. This behavior can be attributed to the special
interplay between phase separation in the directions parallel and
perpendicular to the walls (Figs.~\ref{fig2}-\ref{fig4}). 
Since
the redistribution of both density (Fig.~\ref{fig3}) and relative
concentration (Fig.~\ref{fig4}) between different $z$ is so
pronounced during this range of times, the
time evolution for a given value of $z$ is similar to the time 
evolution of a system whose
order parameter is not conserved.
In the thin film as a
whole, however, both particle numbers are conserved; hence, the density and
concentration are conserved variables, when we
consider the total film. Only for times $t=500$ or larger the
profiles of density $\rho(z)$ and concentration $c(z)$ are practically
independent of time, and then in a particular layer (i.e.,
particular value of $z$) the order parameters behave as if they
were strictly conserved. Gratifyingly, in the time region from
200$\leq t \leq  2000$ the data for $R_1 (z,t)$ and $R_2(z,t)$
show indeed a much more standard behavior, being essentially
independent from the cutoff $q_{\rm cut}$, and almost independent
of $z$, showing that now a well-defined unique length scale exists
in the system. Figs.~\ref{fig7},~\ref{fig8} reveal that in this
range of times we almost find straight lines at the log-log plots,
with a slope slightly below 1/3. For $t \geq 2000$ the curves even
get slightly flatter, so the growth gets slower; we attribute this
effect to the onset of finite size effects. An important finding
of our study, however, is that we do not see any evidence for the
anomalous law $\ell(t) \propto t^{2/3}$ found by Das et al.
\cite{29,30} in a symmetric mixture confined in thin film
geometry. It remains to be understood whether this different
coarsening behavior is primarily due to the lack of symmetry
between phase separating species in our system or due to
different boundary conditions at the walls.

\subsection{Equal-time correlation functions in real space}

In the context of simulations, it has some practical advantages to
extract characteristic lengths from the equal-time correlation
functions in real space \cite{60} rather than using the structure
factors. In fact, also the result for a length $\ell(t)$ growing
as $\ell(t) \propto t^{2/3}$ was extracted from such a real-space
analysis \cite{29,30}. Thus, it is of interest to study real-space
correlations in the present context, too, to check whether the
findings of Figs.~\ref{fig7} and ~\ref{fig8} are corroborated.

We define a normalized pair correlation function $G(r,z,t)$ as
\begin{equation} \label{eq13}
G(r,z,t) = \frac {g(r,z,t)-1 }{ \tilde{g}(r=0,z,t)-1} \quad ,
\end{equation}
\noindent where $g(r,z,t)$ in the equal-time radial distribution function
for effective monomers of hexadecane in a slice $z$ of width $2
\Delta z=1.5 \sigma$ at the time $t$. 
Here, the distance $r=|\vec{r}_k(t) - \vec{r}_\ell (t)|$ and 
the coordinates $z_k(t)$, $z_\ell(t)$ of
the particles labeled as $k$ and $\ell$ are restricted to this
slice, as in Eq.~(\ref{eq9}). 
The value of $\tilde{g}(r=0,z,t)$, which is used to normalize 
$G(r,z,t)$, we obtain by 
extrapolating $g(r,z,t)$ from the region $r>4\sigma$ to $r=0$ as 
described below, thus 
ignoring the local packing effects, which are present in 
the radial distribution function at short distances.

Fig.~\ref{fig9} shows such data for the layer 3 ($z=3.75 \sigma$) (a) and
the layer 5 ($z=6.75 \sigma$) (b). While in the center of the film 
(layer 5) the curves intersect the abscissa, 
and hence one could follow the traditional
method \cite{27,30} to define a characteristic domain linear
dimension $\ell(t)$ from the first zero crossing of $G(r,z,t)$,
this method clearly does not work in slices close to the walls:
e.g., for $z=3.75 \sigma$ and 
time $t=40$ (shown in the inset of Fig.~\ref{fig9}a) we rather see a
continuous decay towards the abscissa with a very flat minimum at 
$r\approx 18\sigma$ instead of a clearly identifiable crossing of the abscissa. 
Thus, we tried
heuristically an alternative way to extract a length $\ell(t)$,
by fitting a straight line to $G(r,z,t)$ in the regime $0.6 \leq
G(r,z,t) \leq 1$. The zero-crossing of these straight lines would
allow to identify a length $\ell(t)$ for all values of $z$.
However, this method also is doubtful, particularly for times $t
\leq 100$, since there the curves for $G(r,z,t)$ show strong
oscillations for small $r$. These oscillations are not due to bad
statistics, but simply reflect the liquid short range order:
oscillatory variation of $g(r,z,t)$ due to the packing of
particles  in the nearest neighbor shell, next nearest neighbor
shell, third nearest neighbor shell, etc., around a particle
\cite{58}. This short range order needs to be disentangled from
the growth of a length scale due to phase separation
(Fig.~\ref{fig10}). Only when $g(r)$ is distinctly nonzero for
$r\geq 4 \sigma$, the growing length scale can be identified; therefore,
very short times (such as $t=20$) obviously must be discarded. 
However, in the regime $r \geq 4 \sigma$ and $t \leq 100$, $ g(r,z,t)$
exhibits also clear curvature, and hence any straight line fit prone to
large systematic errors. Therefore, we choose the ad hoc form
\begin{equation} \label{eq14} 
y(r,z,t)=1 + a(z,t) \exp [-r/\tilde{l}(z,t)] + b(z,t)
\end{equation}
\noindent to smooth our data for $g(r,z,t)$ using $a(z,t)$, $\tilde{l}(z,t)$, 
and $b(z,t)$ as fit parameters; $b(z,t)$ is negative when a zero crossing 
(i.e., $g(r,z,t)-1=0$ at $r>4\sigma$) occurs and is positive otherwise. 
This fit-function is used in the range $r>4\sigma$ and $g(r,z,t)>1.01$ 
but actually provides a good representation of the actual data down 
to and below the first-zero crossing for those values 
of $z$ and $t$ where such a zero crossing occurs. 
Also, this function is used to extrapolated from the region $r>4\sigma$ to 
$r=0$ to obtain the normalization constant 
$\tilde{g}(r=0,z,t)=1 + a(z,t) + b(z,t)$ for Eq.~(\ref{eq13}).  
From the first-order Taylor expansion of Eq.~(\ref{eq14}) we obtain 
an approximation to $G(r,z,t)$
\begin{equation} \label{eq15} 
G(r,z,t)\approx \frac{ a(z,t) [1-r/\tilde{l}(z,t)] + b(z,t) }{a(z,t) + b(z,t)}= 1-r/l(z,t) 
\end{equation}
\noindent from which we can also define a characteristic length  
$l(z,t)=\tilde{l}(z,t)(a(z,t) + b(z,t))/a(z,t)$ as a value of $r$ 
at the intersection of a line given by Eq.~(\ref{eq15}) with the axis $G=0$.

Fig.~\ref{fig11} shows the time evolution of the
characteristic length $\ell(z,t)$ extracted in this way for three
choices of $z$. While for $t < 100$, where $\ell (z,t)$ is only of
the order of a few $\sigma$, indeed the data seem to be compatible
with a behavior $\ell(z,t) \propto t^{2/3}$ as observed by Das et
al. \cite{29}, for the decade $100 \leq t \leq 1000$ the data seem
to be compatible with $\ell (z,t) \propto t^{1/2}$  showing a 
crossover to $\ell (z,t) \propto t^{1/3}$ at later time. 
For $t \geq 1000$, a crossover to a still slower growth is evident 
in our simulations, which is however strongly affected by finite 
size effects, since the number of growing (cylinder-shaped) domains 
is already rather small. 
Note that this implies that finite size effects already set in for 
$\ell(z,t)\approx 20 \sigma$, so despite our large system 
($300\sigma\times 300\sigma$) we cannot follow the kinetics of 
spinodal decomposition for a large enough range of times in order to 
make significant statements on the asymptotic power law growth! Much 
larger systems need to be simulated for this purpose.

\section{CONCLUSIONS}

In this work, we have presented computer simulations of spinodal
decomposition of a coarse-grained model for a compressible binary
fluid mixture, which roughly describes hexadecane dissolved in
supercritical carbon dioxide. This system has a very asymmetric phase diagram in
the plane of variables pressure and molar fraction
(Fig.~\ref{fig1}), and the pressure-jump considered in the present
work is strongly off-critical: the number of CO$_2$ molecules is
88400 while the number of C$_{16}$H$_{34}$ chains is 58800
(leading to 294000 effective segments). We find that the phase
separation is a two-step process: in the first step, there is a
strong segregation between solvent and polymer leading to a
layered structure, with solvent rich layers adjacent to the walls,
and a polymer-rich ultrathin polymer film ``sandwiched'' in
between. This stratified structure, however, is unstable: the
free-standing polymer film in the center of the slit pore breaks
up, CO$_2$-rich bubbles form, and finally a pattern develops with
cylinder-shaped CO$_2$-rich domains, the radius of which grows
with a $\ell(t) \propto t^{1/3}$ law (at the latest stages
accessible to our simulation, as long as finite size effects are
still negligible). For the earlier stages of phase separation,
where a strong coupling between the phase separation in
perpendicular and parallel directions (with respect to the walls)
occurs, we conclude that a description of the structure in terms
of power laws of characteristic linear dimensions is somewhat
misleading, since characteristic lengths extracted from the
structure factor and from the pair correlation function are quite
incompatible with each other. We suggest that there is no simple
scaling behavior in this regime. Note that although we have chosen
the potential between wall particles and solvent particles
identical to the potential between wall particles and effective
monomers, both in the initial and final stages of phase separation
there is significant enrichment of CO$_2$ near the walls, although
our snapshot pictures (resolved as function of $z$ and $t$ in
Fig.~\ref{fig2}) indicated that the data still belong to an
incomplete wetting regime.

Thus, it would be interesting to have a more detailed theoretical
understanding from analytical theory for phase separation with two
coupled order parameters (density and concentration, in our case).
Additional simulations would be valuable where wall-particle
interactions are chosen such that a strictly ``neutral wall''
situation is achieved, where no surface enrichment occurs, and
hence the pure confinement effect on phase separation (not
disturbed by the formation of precursors of wetting layers) could
be studied. 
Also, it would be clearly worthwhile to study more systematically how the phase 
separation kinetics depends on slit thickness, composition of the mixture,
and quench depth. We did some preliminary runs at one different
quench depth in which the system volume was increased by factor 1.73 instead of 1.95 
discussed in our paper and found a rather similar behavior. 
However, significantly different behavior is expected for shallow quenches through 
the critical point, since the correlation length of density and concentration fluctuations 
can exceed the slit thickness, and formation of a stratified structure is not expected.
Finally, in order to get rid of finite size effects,
simulations with billions of particles on massively parallel
supercomputers would be required. 
However, such  detailed and computationaly extensive studies 
would be a very challenging task for presently available computer resources and 
itself could be a topic for forecoming papers.
We also hope the present work
will stimulate more work on the kinetics of phase
separation in nanoscopic confinement such as very thin channels.\\

{\bf Acknowledgements}: One of us (K. Bucior) acknowledges support
via an Alexander von Humboldt-Fellowship. We are grateful to the
J. von Neumann Institute for Computing (NIC) for access to the
JUMP multiprocessor and to the EU Network of Excellence for access 
to the SOFTCOMP Cluster at J\"ulich Supercomputing Centre (JSC). 
Helpful discussions with S.K.
Das, J. Horbach, M. M\"uller, W. Paul and P. Virnau are
acknowledged. We also thank T. St\"uhn for his advice with the
ESPResSo software.

\bigskip

\clearpage

\clearpage

\section*{FIGURE CAPTION}

FIG.1. Isothermal slice through the binary phase
diagram of the present model for CO$_2$ + C$_{16}$H$_{34}$
mixtures at $T=486.2$K (reduced temperature $T^*=1.16$) in the bulk, 
using the pressure $p$ 
and the molar fraction $x$ of CO$_2$ as variables. The
coexistence curve encloses a two-phase coexistence region
containing a polymer-rich phase (left) and a supercritical CO$_2$
vapor (near $x=1$, right). The simulations of quenching
experiments discussed in the present paper are done for $x=0.6$.
This phase diagram is taken from the results of Ref. \cite{35}.\\

FIG.2. Snapshot picture showing the structure
formation after the quench for $L_x \times L_y=300\sigma$ slices of width
$1.5\sigma$ centered at $z=3.75 \sigma$ (a-e) and at $z=6.75
\sigma$ (f-j). 
Snapshots are presented at times 
$t=10$ (a, f), $50$ (b, g), $100$ (c, h), $200$ (d, i) and $1000$ (e, j). 
The insets in snapshots (c) and (h) illustrate the enlarged regions in 
the left-bottom corner of size $30\sigma \times 30\sigma$ (marked by rectangles): 
the gray spheres correspond to the supercritical solvent molecules, 
and the black ones represent the chain molecules.\\

FIG.3. Total density profile $\rho_{\rm tot} (z)$
(laterally averaged in the thin film)
plotted vs. position across the slit pore  
$z/L_z$ for the initial state before the quench 
($L_z=12\sigma$, $L_x=L_y=240\sigma$),
and for different stages of phase separation 
($L_z=15\sigma$, $L_x=L_y=300\sigma$). 
The curves are shown at different times after the quench, as indicated.
\\

FIG.4. Relative concentration profile $c(z)$ of the solvent
laterally averaged in the thin film
plotted vs. position across the slit pore $z/L_z$ 
for the same times and the system dimensions as 
shown in Fig.~\ref{fig3}.
\\

FIG.5. Time evolution of the reduced temperature
$T^*$ (top) and the reduced pressure $p^*$ (bottom) before and after the quench.
The time of the quench is at $t=0$.
\\

FIG.6. Density-density structure factor $S_{\rm nn}
(q,z,t)$ in the initial state after equilibration (just before the
quench) for the system of the linear dimensions $L_x=L_y=240 \sigma$, 
$L_z= 12 \sigma$ (a), and after
the quench at times $t=40$ (b) and $t=100$ (c) for the system dimensions
$L_x=L_y=300 \sigma$, $L_z=15 \sigma$. Various choices of
$z$ are included, as indicated in the figure. In case (a), an
average over 130 configurations was performed, while in cases (b)
and (c), 30 configurations were averaged over. 
Note: the two curves for layers 5 and 6 (in the middle of the slit pore) 
cannot be distinguished 
on the scale shown in figure (c). In all cases, the film was divided 
into 10 slices of 
width
$2 \Delta z = L_z/10$.
\\

FIG.7. Log-log plot of the characteristic domain
size $R_1(z,t)$ (calculated from the density-density structure 
factor  $S_{\rm nn}$ shown in Fig.~\ref{fig6}) 
vs. time 
for four choices of $z$, as indicated in the figure, 
and three plausible choices of the cutoff for the  
wave vector: $q_{\rm cut} =1,2$, or $3$, respectively.
\\

FIG.8. Log-log plot of the characteristic domain
size $R_2(z,t)$ (calculated from the concentration-concentration 
structure factor $S_{\rm cc}$) vs. time 
for four choices of $z$, as indicated
in the figure, and three plausible choices of the cutoff for the 
wave vector: $q_{\rm cut} =1,2$, or $3$, respectively.
\\

FIG.9. The real-space normalized pair 
correlation function $G(r,z,t)$ [Eq.~(\ref{eq13})] vs.
distance $r$ for $z=3.75 \sigma$ (a) and $z=6.75 \sigma$ (b) 
at various times, as indicated in the figure. 
The inset in part (a) shows a magnification of the region 
near the origin for early stage of spinodal decomposition. 
Note, that the curve for $t=40$ exhibits a very flat local 
minimum near $r\approx 18\sigma$, which prohibits using the 
first-zero crossing of $G(r,z,t)$ to measure the characteristic 
domain size in this range of the time. Straight lines 
illustrate fits to estimate the characteristic domain length 
scale $l(z,t)$ from an effective initial slope of these 
curves at $r=0$ using Eq.~(\ref{eq15}). 
\\

FIG.10. Plot of the radial distribution function $g(r,z,t)$ 
for the middle layer ($z=6.75 \sigma$) vs. distance $r$ 
for various times, as indicated.
\\

FIG.11. Log-log plot of characteristic domain
linear dimension vs. time calculated in three layers of
thickness $\Delta z =1.5 \sigma$ parallel to the walls, centered
at $z=3.75 \sigma$, $5.25 \sigma$ and $z=6.75 \sigma$,
respectively. The domain size shown here is defined in the text 
by Eq.~(\ref{eq15}). Straight lines are the guide for eyes
illustrating the power laws $\ell(t) \propto t^{1/3}$ (dashed line),
$\ell (t) \propto t^{1/2}$ (dashed-dotted line) and $\ell(t)
\propto t^{2/3}$ (dashed-double dotted line), respectively.


\clearpage

\begin{figure}[h]
\begin{center}
\includegraphics[width=8cm,clip]{Fig_1.eps}
\caption{\label{fig1} K. Bucior}
\end{center}
\end{figure}

\clearpage

\begin{figure}
\begin{center}
\includegraphics[width=8cm,clip,angle=-0]{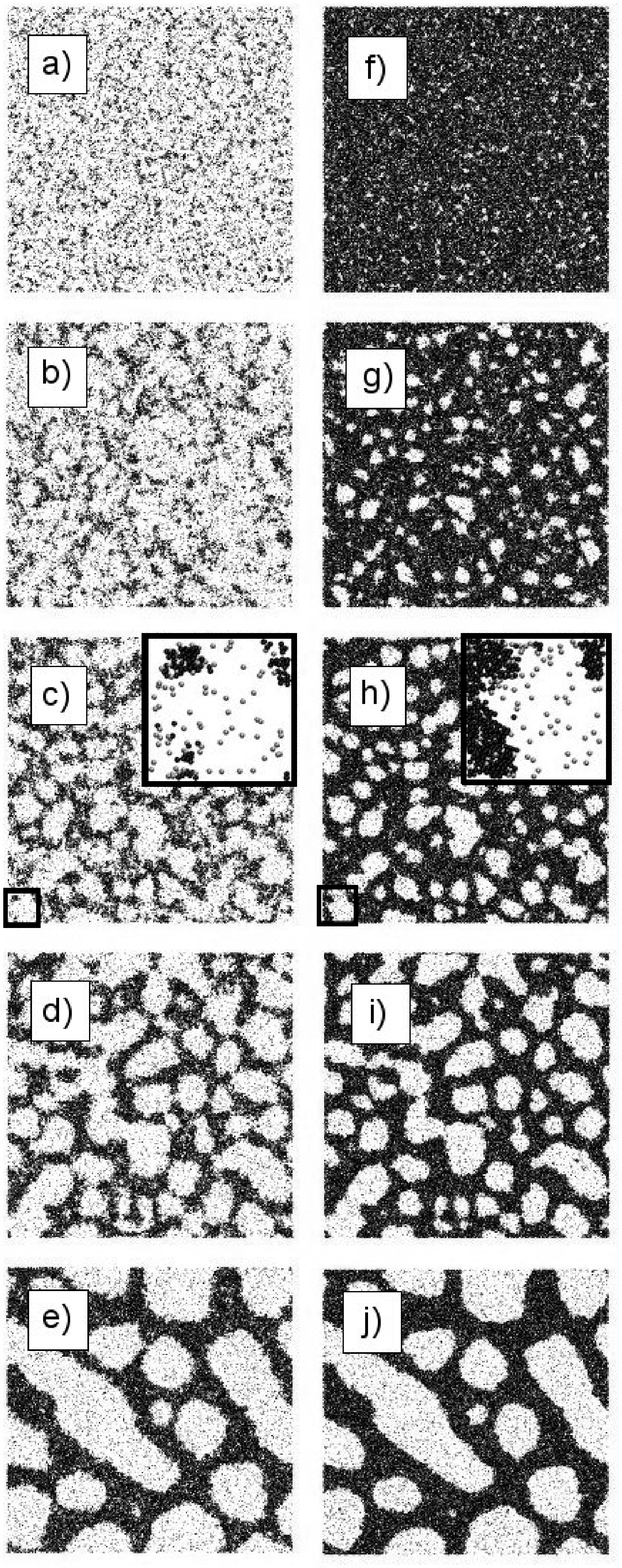}
\caption{\label{fig2} K. Bucior}
\end{center}
\end{figure}

\clearpage

\begin{figure}
\begin{center}
\includegraphics[width=8cm,clip,angle=-0]{Fig_3.eps}
\caption{\label{fig3} K. Bucior}
\end{center}
\end{figure}

\clearpage

\begin{figure}
\begin{center}
\includegraphics[width=8cm,clip,angle=-0]{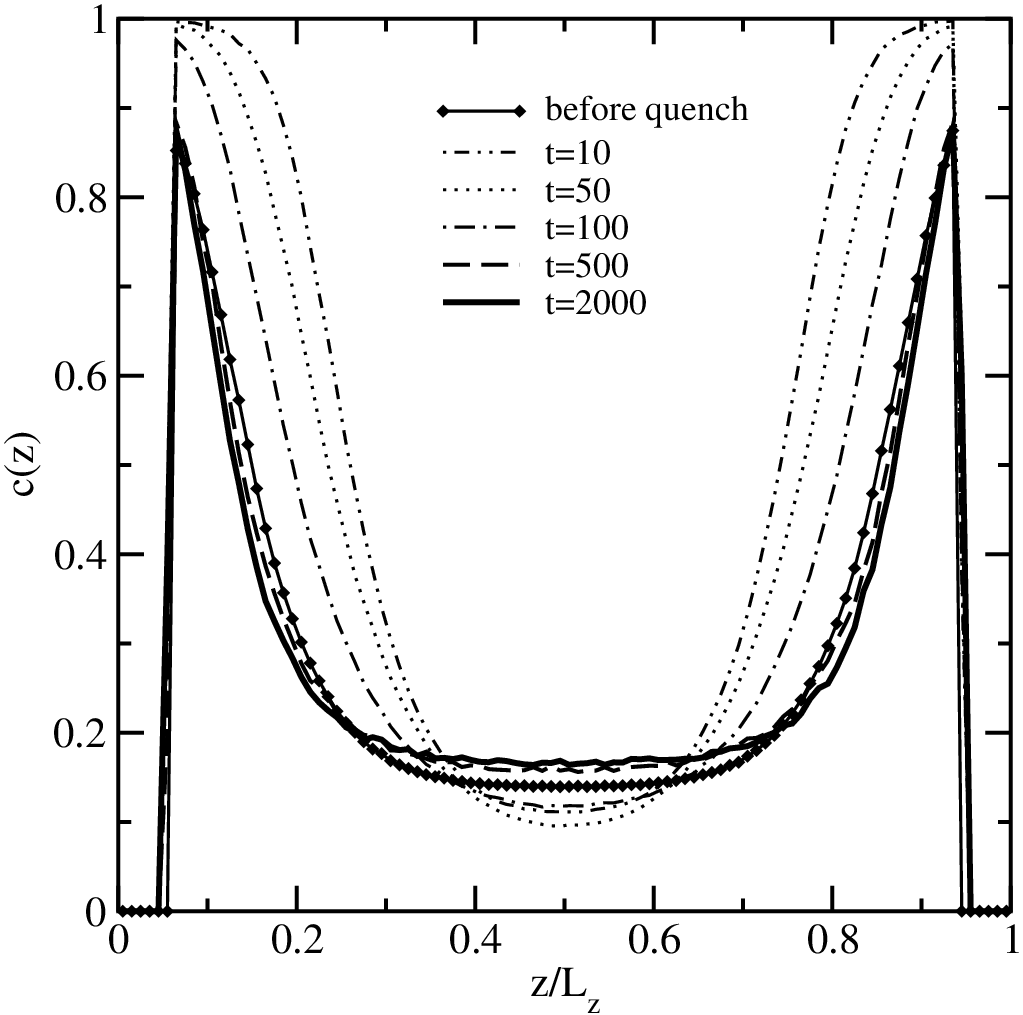}
\caption{\label{fig4} K. Bucior}
\end{center}
\end{figure}

\clearpage

\begin{figure}
\begin{center}
\includegraphics[width=8cm,clip,angle=-0]{Fig_5.eps}
\caption{\label{fig5} K. Bucior}
\end{center}
\end{figure}

\clearpage

\begin{figure}
\begin{center}
\includegraphics[width=7.5cm,clip,angle=-0]{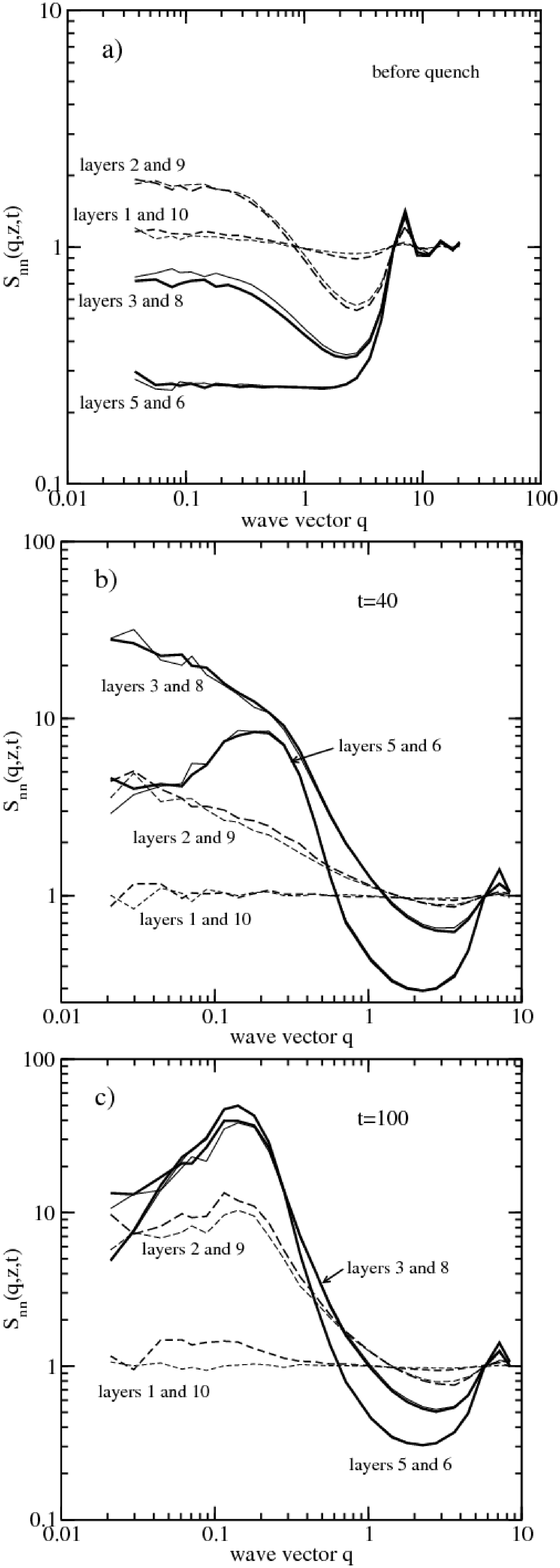}
\caption{\label{fig6} K. Bucior}
\end{center}
\end{figure}

\clearpage

\begin{figure}
\begin{center}
\includegraphics[width=8cm,clip,angle=-0]{Fig_7.eps}
\caption{\label{fig7} K. Bucior}
\end{center}
\end{figure}

\clearpage

\begin{figure}
\begin{center}
\includegraphics[width=8cm,clip,angle=-0]{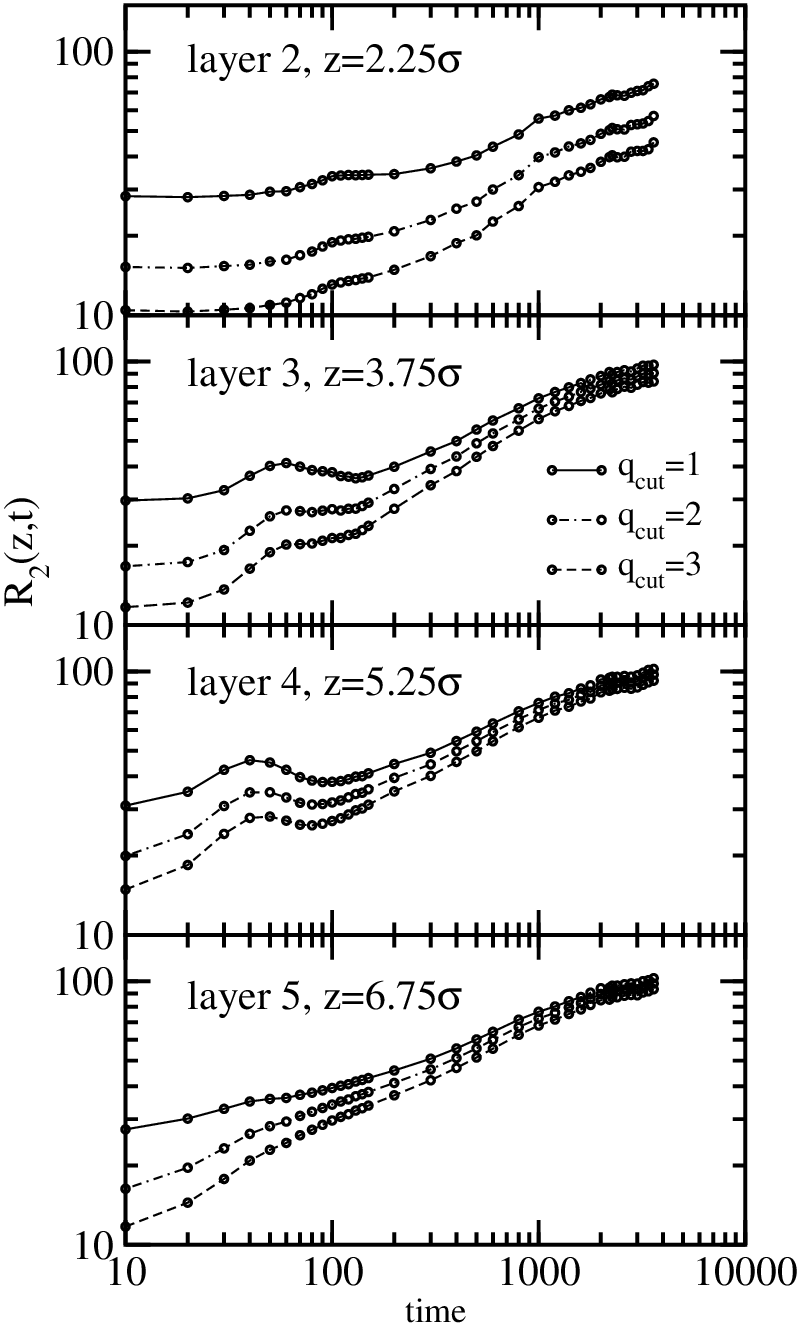}
\caption{\label{fig8} K. Bucior}
\end{center}
\end{figure}

\clearpage

\begin{figure}
\begin{center}
\includegraphics[width=8cm,clip,angle=-0]{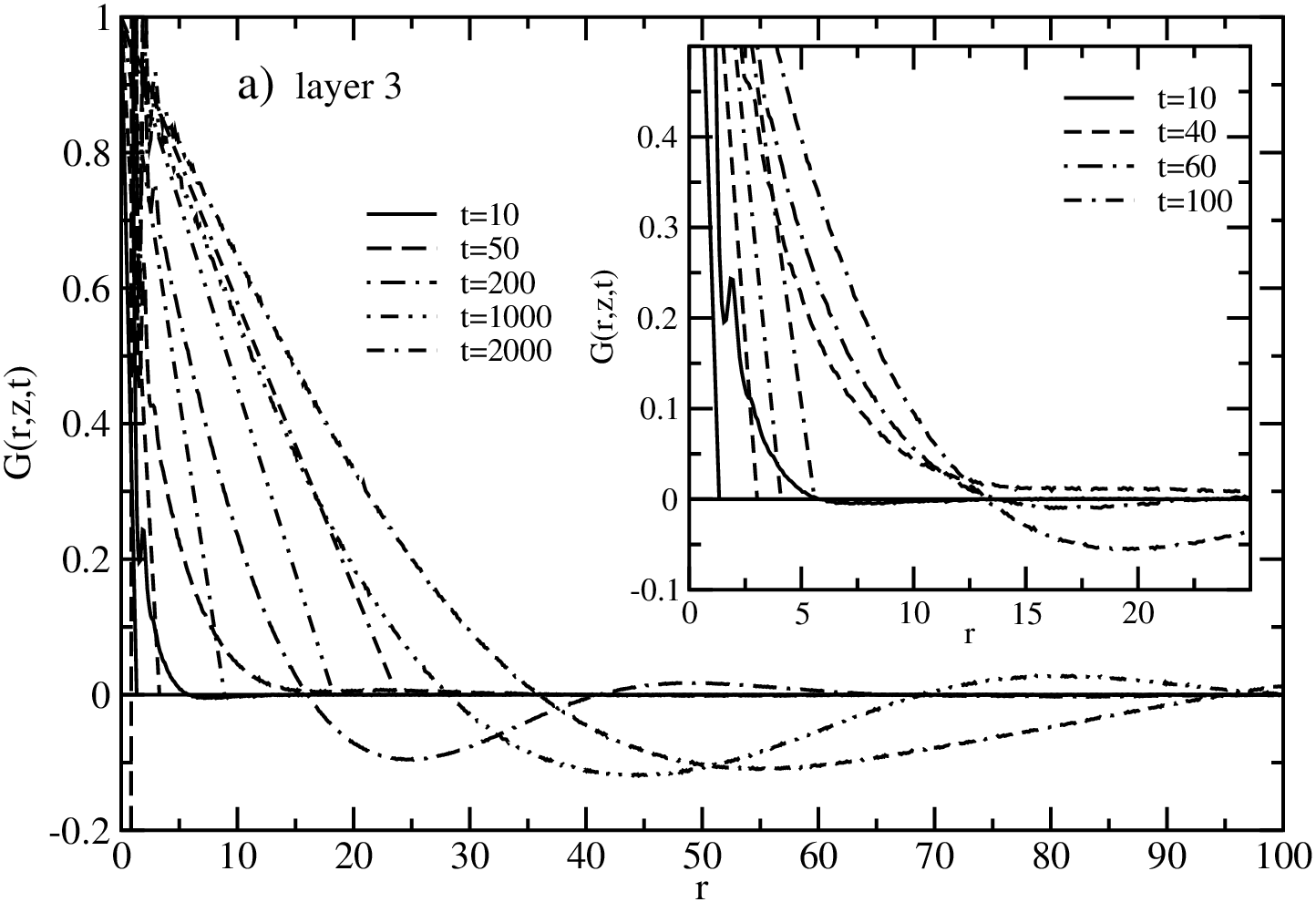}
\includegraphics[width=8cm,clip,angle=-0]{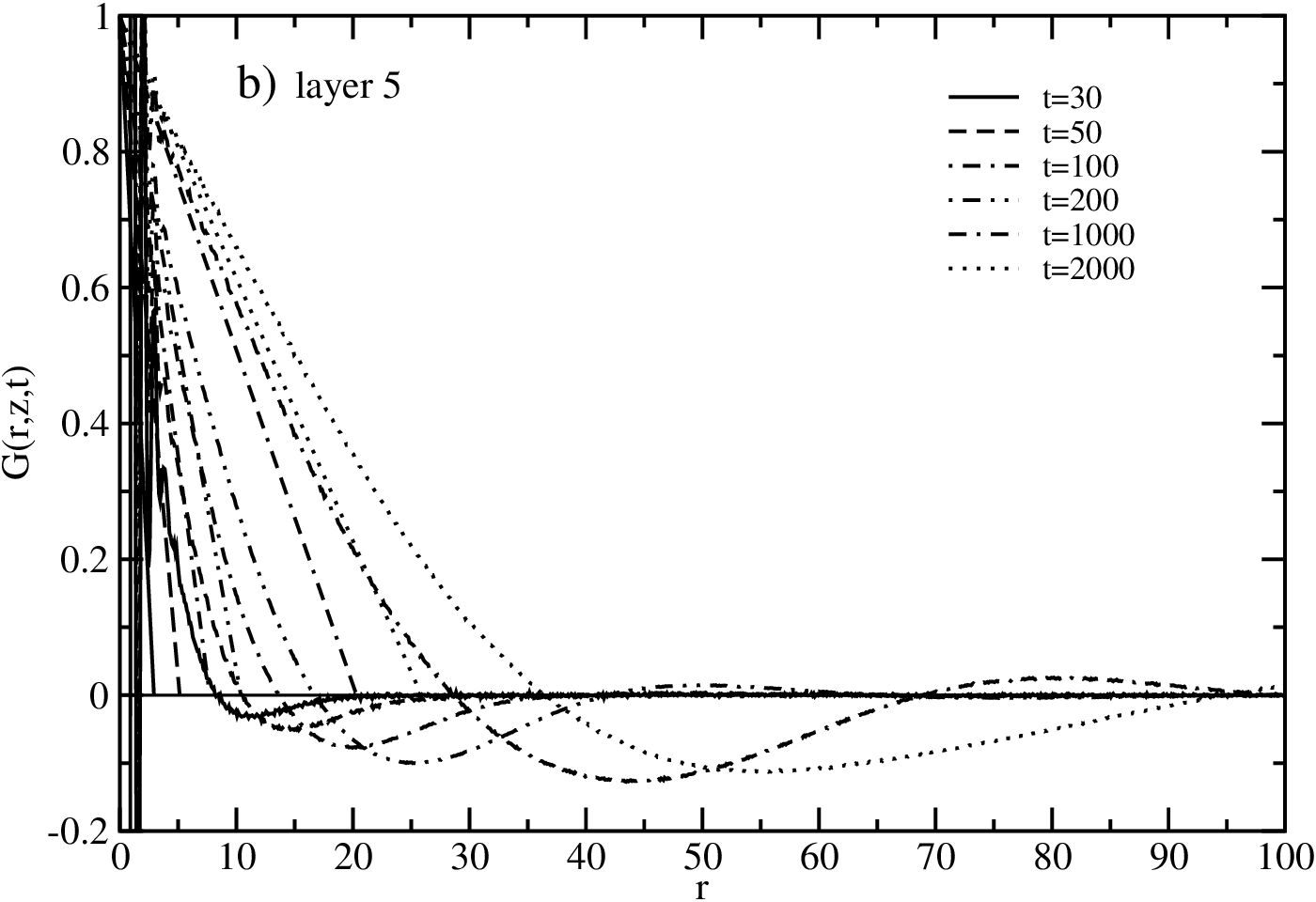}
\caption{\label{fig9} K. Bucior}
\end{center}
\end{figure}

\clearpage

\begin{figure}
\begin{center}
\includegraphics[width=8cm,clip,angle=-0]{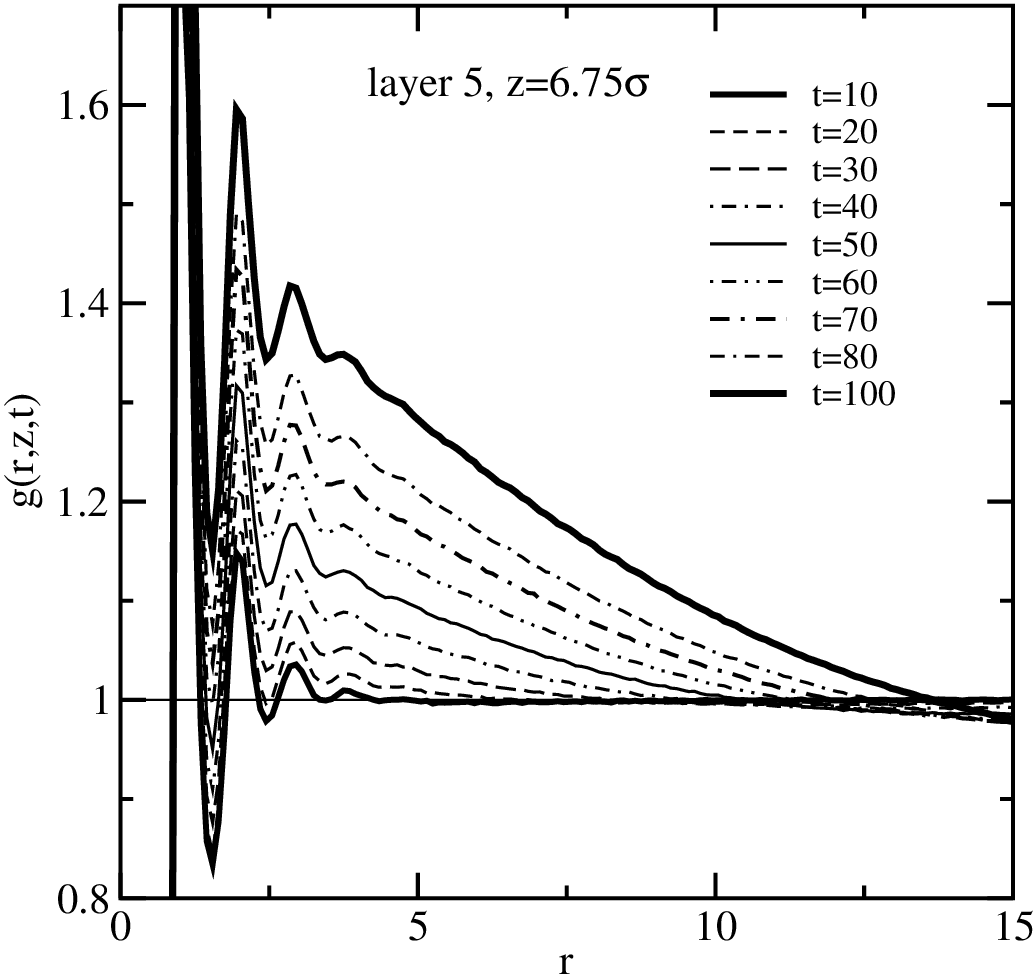}
\caption{\label{fig10} K. Bucior}
\end{center}
\end{figure}

\clearpage

\begin{figure}
\begin{center}
\includegraphics[width=8cm,clip,angle=-0]{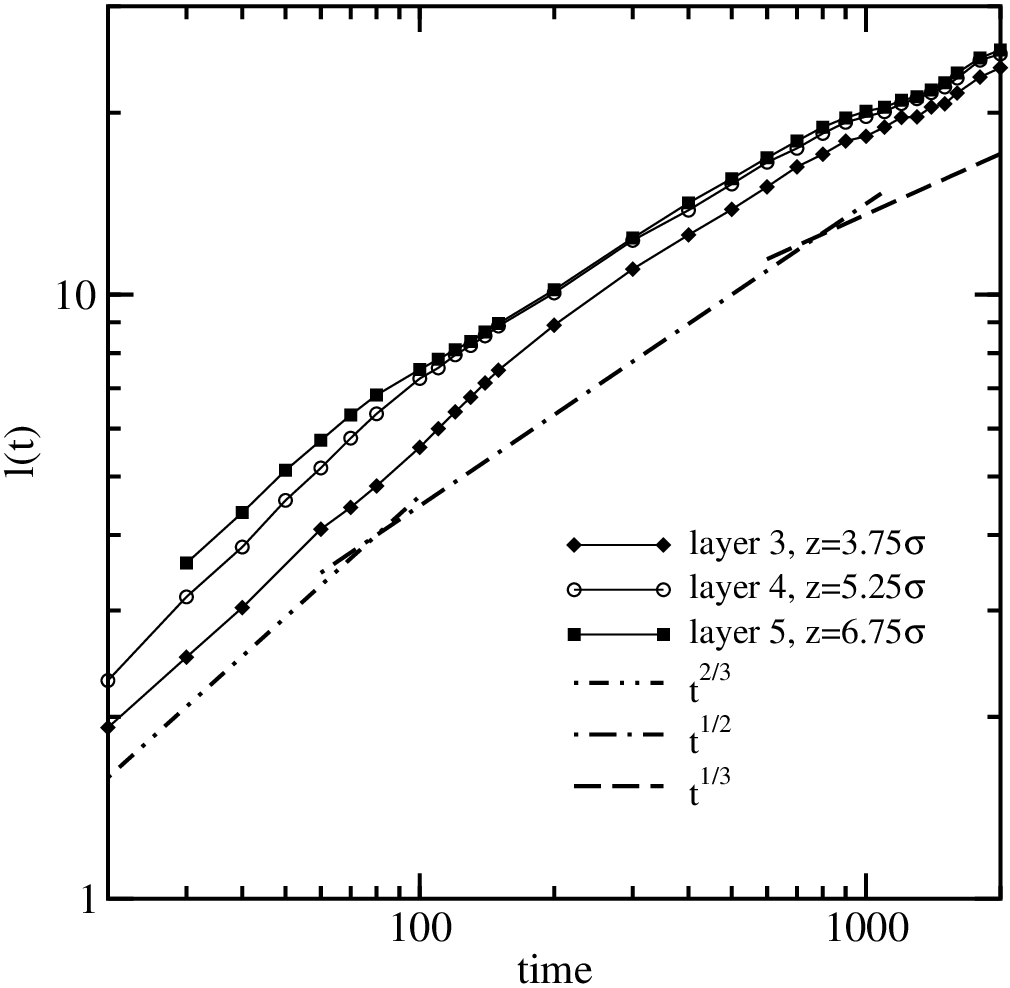}
\caption{\label{fig11} K. Bucior}
\end{center}
\end{figure}

\end{document}